\documentclass[prc,aps,preprint,showpacs,amssymb,superscriptaddress,endfloats]{revtex4}

\usepackage{amsmath}
\usepackage{graphicx}
\usepackage{dcolumn}
\usepackage{float}

\begin{document}

\title{Shell-model study of single-neutron strength fragmentation in $^{137}$Xe}

\author{L. Coraggio}
\affiliation{Istituto Nazionale di Fisica Nucleare, \\
Complesso Universitario di Monte  S. Angelo, Via Cintia - I-80126 Napoli,
Italy}
\author{A. Covello}
\affiliation{Istituto Nazionale di Fisica Nucleare, \\
Complesso Universitario di Monte  S. Angelo, Via Cintia - I-80126 Napoli,
Italy}
\affiliation{Dipartimento di Scienze Fisiche, Universit\`a
di Napoli Federico II, \\
Complesso Universitario di Monte  S. Angelo, Via Cintia - I-80126 Napoli,
Italy}
\author{A. Gargano}
\affiliation{Istituto Nazionale di Fisica Nucleare, \\
Complesso Universitario di Monte  S. Angelo, Via Cintia - I-80126 Napoli,
Italy}
\author{N. Itaco}
\affiliation{Istituto Nazionale di Fisica Nucleare, \\
Complesso Universitario di Monte  S. Angelo, Via Cintia - I-80126 Napoli,
Italy}
\affiliation{Dipartimento di Scienze Fisiche, Universit\`a
di Napoli Federico II, \\
Complesso Universitario di Monte  S. Angelo, Via Cintia - I-80126 Napoli,
Italy}

\date{\today}

\begin{abstract}
We have performed shell-model calculations for the nucleus $^{137}$Xe, which was recently
studied experimentally using the $^{136}$Xe($d,p$) reaction in inverse kinematics.  The main aim of our study has been to investigate the single-neutron properties of the observed states, focusing attention on the spectroscopic factors.
We have employed a realistic low-momentum two-body effective interaction derived 
from the CD-Bonn nucleon-nucleon potential that has already proved quite successful 
in describing the spectroscopic properties of nuclei in the $^{132}$Sn region.
Comparison shows that our calculations  reproduce very well the experimental excitation
energies and yield spectroscopic factors that come close to those extracted from the data.

\end{abstract} 


\maketitle

In a recent paper \cite{Kay11}, the single-neutron structure of the nucleus 
$^{137}$Xe was investigated at Argonne National Laboratory via the $^{136}$Xe ({\it d,p}) reaction 
in inverse kinematics. In particular, spectroscopic factors for several states in $^{137}$Xe were extracted from the data by DWBA calculations, providing valuable information on 
the evolution of single-neutron states outside  $^{132}$Sn, which is a subject of great 
current interest \cite{Jones11,Schiffer12}.

Actually, the ({\it d,p}) reaction on $^{136}$Xe was first studied more than 40 years ago in normal kinematics \cite{Schneid66,Moore68}. Some twenty years later, a new study of this reaction was performed at GSI in inverse kinematics
 \cite{Kraus91} to test the effectiveness of this method as a tool for the study of nuclei far from stability with transfer reactions. 
This was  also a main motivation for the experiment of \cite{Kay11}, which provided a stringent test of the HELIOS spectrometer used to analyze the outgoing protons. In this experiment the beam of $^{136}$Xe was delivered at an  energy about twice as high as that used in  \cite{Kraus91}, which implies larger cross sections for transfer to high-$l$ states. 

In other words, the contribution of \cite{Kay11} is two-fold. On the one hand it adds to our understanding of the single-particle structure of  $^{137}$Xe. On the other hand, it helps pave the way to perform transfer reactions with highly unstable nuclei, which is a major goal of the next generation of radioactive ion beam facilities.
This has stimulated the present study, which aims both to interpret the  data available for $^{137}$Xe
and encourage further experimental efforts.

Over the past several years we have performed various shell-model studies 
\cite{Coraggio09,Covello11} of neutron-rich nuclei beyond $^{132}$Sn, which have all yielded results in very good agreement with experiment. In these studies, a unique Hamiltonian has been used with the single-particle energies taken from experiment and the two-body effective interaction derived from the CD-Bonn nucleon-nucleon ($NN$) potential ~\cite{Machleidt01} without using any adjustable parameter.

The new findings mentioned above have led us us to make use of this Hamiltonian to investigate the single-neutron properties of $^{137}$Xe. It should be  mentioned here that in  a previous work  \cite{Liu10} we have performed a study of this nucleus focusing attention on high-spin states produced in the spontaneous fission of $^{252}$Cf.

In our calculations, we assume that the four valence protons outside doubly magic $^{132}$Sn occupy the five orbits $0g_{7/2}$, $1d_{5/2}$, $1d_{3/2}$, $0h_{11/2}$, and $2s_{1/2}$ of the 50-82 shell
while the odd neutron has available the six orbits $1f_{7/2}$, $2p_{3/2}$, $2p_{1/2}$, $0h_{9/2}$,
$1f_{5/2}$, and  $0i_{13/2}$ of the 82-126 shell. 
We take as single-proton and -neutron energies the experimental energies of the levels with corresponding spin and parity in $^{133}$Sb and $^{133}$Sn \cite{ENSDF}, respectively.
There are, however, two exceptions since no $1/2^-$ state in $^{133}$Sb and  $13/2^+$ state in $^{133}$Sn has yet been observed. Therefore, we take the proton $2s_{1/2}$ and the neutron $0i_{13/2}$ energies from Refs. \cite{Andreozzi97,Coraggio02}, respectively,  where it is discussed how they are determined.
It should be noted that we place the neutron $2p_{1/2}$ level at 1.363 MeV excitation energy, which is about 300 keV lower than the previous proposed value \cite{Hoff96}. This new value comes from the recent work of Ref. \cite{Jones10}, where the short-lived $^{133}$Sn has been studied 
by performing the  $^{132}$Sn({\it d,p}) reaction in inverse kinematics. It is worth pointing out here that this experiment has provided clear evidence for the purity of the $7/2^-$, $3/2^-$, $1/2^-$ and $5/2^-$ single-neutrons states. 			

The two-body  effective interaction has been derived within the framework of
perturbation theory~\cite{Coraggio09b,Coraggio12} starting, as mentioned before,  from the 
CD-Bonn $NN$ potential renormalized by way of the $V_{\rm low-k}$ 
approach~\cite{Bogner02}.  
More precisely, we start by  deriving $V_{\rm low-k}$
with a cutoff  momentum $\Lambda=2.2$ fm$^{-1}$. Then, using this  
potential plus the Coulomb force for protons, we calculate  the matrix elements of 
the effective interaction  by means of the $\hat Q$-box folded-diagram expansion, with 
the $\hat Q$-box  including all diagrams up to second order  in the 
interaction.
These diagrams are computed within the harmonic-oscillator basis using intermediate states 
composed of all possible hole states and particle states restricted to the five proton and 
neutron shells above the Fermi surface. The oscillator parameter is 7.88 MeV, 
as obtained from the expression  $\hbar \omega= 45 A^{-1/3} -25 A^{-2/3}$ with $A=132$.
The calculations have been performed by using the OXBASH code \cite{OXBASH}.

 In Table I we report the calculated excitation energies and spectroscopic factors and compare them with those obtained from the ($d,p$) experiment of Ref.
\cite{Kay11}. We only consider experimental states up to the $13/2^+$ one at 1.751 MeV. In the higher-energy region we did not make any attempt to identify the 
observed states with the calculated ones. In fact, the available experimental 
information  is very scanty while a large number of states with spin and parity 
corresponding to those of the single-neutron orbits are predicted by the 
theory. It should be mentioned that  all the states reported in Table I were previously known~\cite{ENSDF}, except the second $9/2^-$ state at 1590 keV and the $13/2^+$ state at 1751 keV,
the spin-parity assignment to the latter being in accord with a previously unpublished
assignment, as mentioned in \cite{Kay11}. Note that the two 
states at 0.986 and 1.534 MeV reported in \cite{Kay11} with  ambiguous spin assignments have been identified as $1/2^-$ and $5/2^-$ states in other experiments~\cite{ENSDF}. In this regard, it is worth remarking that they are likely to correspond to the $1/2^-$ and $5/2^-$ states at  0.91 and 1.41 MeV observed in  \cite{Moore68}. The discrepancies between the excitation energies measured in the two experiments may be due the low precision of energy determination in the earlier one \cite{ENSDF}.

The experimental energies are well reproduced by the theory, with
discrepancies not exceeding 150 keV, the only exception being  the $13/2^+$ state, whose  energy is overestimated by about 330 keV. This state is found to be essentially of single-particle nature and therefore its energy is 
quite sensitive to the position of the $0i_{13/2}$ neutron orbit, 
which, as mentioned above, is not experimentally known. 
Our calculations confirm the $1/2^-$ assignment for the state at 0.986 MeV,  the second $3/2^-$ state being predicted at a much higher energy
(about 1.8 MeV).  As for the level at  1.534 MeV,  we cannot discriminate between 
$J^{\pi}=5/2^-$  and  $7/2^-$, the calculated states lying  both close to this energy. We will come back to this point later.
  
Let us now turn to the spectroscopic factors. In our discussion, we find it 
appropriate to consider also the early results of Ref. \cite{Moore68}. In fact,
 in that work the spectroscopic factors included  in Table I, but those of 
the $13/2^+$ and the second $9/2^-$ states, are reported and their  values are
 on the whole not very different from those of \cite{Kay11}.  
The main differences  relate to the spectroscopic factors of the ground  $7/2^-$ and   yrast $9/2^-$ states, the values of \cite{Kay11} (see Table I) being substantially larger than those of \cite{Moore68}, 0.68 and 0.31, respectively.

As can be seen  in Table I, an overall good agreement is found between the calculated spectroscopic factors and those obtained in  Ref. \cite{Kay11}, most of 
the theoretical values differing from the experimental ones by no more than 23\%. In particular, our value for the ground state, 0.86, gives support 
to that obtained in \cite{Kay11}. However, the calculated spectroscopic factor 
of the yrast $9/2^-$ state largely overestimates
the value of  both \cite{Moore68} and \cite{Kay11}. As a matter of fact, we underestimate the  fragmentation  of the $9/2^-$ strength. We predict, indeed,  a second $9/2^-$ state with essentially no single-particle strength, its spectroscopic factor being  0.01, to be compared with the experimental value of 0.24 \cite{Kay11}. It should be noted, however, that no evidence for this state was found in \cite{Moore68}. 

The weak fragmentation of the $9/2^-$ strength resulting from our calculations 
may be due to a somewhat too strong neutron-proton interaction. In fact, if one
 ignores  the neutron-proton interaction, then  the single-particle strength is completely concentrated in the second $9/2^-$ state, lying 250 keV above the yrast one. This may be easily seen when writing the first and second $9/2^-$ unperturbed states as $|^{136}{\rm Xe} \ {\rm g.s.;} \ \nu h_{9/2}>$ and   $|^{136}{\rm Xe} \  2^{+}{\rm}; \ \nu f_{7/2}>$, respectively.

The spectroscopic factor of the $5/2^{-}, 7/2^{-}$ level ( identified as 
$5/2^-$ in \cite{Moore68} with $C^{2}S=0.16$) is underestimated by the theory, 
either if this level is associated  with the calculated $7/2^-$ or the $5/2^-$ state. However, based on the small spacing, about 75 keV,  predicted between these two states, one cannot exclude that the peak observed at 1.534 MeV, as well as  that at 1.41 MeV in \cite{Moore68}, consists of two unresolved levels, the extracted spectroscopic factor corresponding to the sum of their spectroscopic factors.

As for the $13/2^+$ states, only the yrast one has been observed in \cite{Kay11}.
However, based on the work of Ref. \cite{Kay08}, the authors of \cite{Kay11} give an estimate  of the energy and spectroscopic factor of the second  $13/2^+$ state, 3360(110) keV and  0.15(4), respectively, which is consistent with a $13/2^+$ assignment to either of the two 
observed peaks at 3.310 and 3.470 MeV. We predict a 
second $13/2^+$ state at 3.452 MeV with a spectroscopic factor of 0.01.

As a general comment, we may say that the predicted single-particle strength, in agreement with experiment, is strongly concentrated in the yrast states only for $J^{\pi}=7/2^-$ and $13/2^+$.
This is not the case for the $J^{\pi}=1/2^-$, $3/2^-$  and $5/2^-$ states. However, we find that for the first two angular momenta the spectroscopic factors of the yrast states are still the largest  ones with the single-particle strength mainly distributed over the lowest-lying states. In fact, it is sufficient to sum up to an energy of 2 MeV, corresponding to the first  three and four lowest-lying states, respectively, to obtain more than 75\% of the strength. The situation is quite different for  $J^{\pi}=5/2^-$.  In this case, by summing over the first 20 states, namely up to 3 MeV, we get only 65\% of the strength, the largest spectroscopic factor 
being 0.20 for the 5th excited $5/2^-$ state at 2.039 MeV.

It is now interesting to discuss our results for the centroids
of the $0i_{13/2}$ and $0h_{9/2}$ neutron orbits.
We find that the energy difference between these two centroids is 0.8 MeV,  with a decrease
of about 0.3 with respect to the  $^{133}$Sn initial value, which is in 
agreement with the results reported in \cite{Kay11}. To understand the reason for the reduction in the separation of these two orbits, it is convenient to write \cite{Umeya06,Duguet12} the energy centroid as

\begin{equation}
\bar{\epsilon}_{j_{\nu}}=\epsilon_{j_{\nu}} + \sum_{j_{\pi}} 
V^{M}(j_{\nu} j_{\pi}) N_{j_{\pi}},
\label{one}
\end{equation}

\noindent {where $\epsilon_{j_{\nu}}$ is  the energy of level $j_{\nu}$
in  $^{133}$Sn,
 $N_{j_{\pi}}$ denotes the number of protons occupying the $j_{\pi}$ orbit in the even-even system  and 
$V^{M}(j_{\nu}j_{\pi})$ the monopole component of the interaction.}

By examining the various terms of Eq. (1), it may be seen  that the lowering of the $0i_{13/2}$ with respect to the $0h_{9/2}$ orbit  stems from the fact that the value of  
$V^{M}({\nu}i_{13/2}{\pi}g_{7/2})$, -0.33 MeV, is larger than that of 
$V^{M}({\nu}h_{9/2}{\pi}g_{7/2})$ by a factor of about 2. In fact, the 
components which play a dominant role in the sum are only those involving the 
$\pi g_{7/2}$ orbit, owing to the large occupancy of the latter, 3.1.
The remaining proton occupancy, 0.9, is distributed between  the $1d_{5/2}$ and $0h_{11/2}$ orbits and  does not make a significant contribution. It is worth mentioning that our calculated proton occupancies for
the ground state of  $^{136}$Xe are in very good agreement with those obtained from pick-up and 
stripping reactions \cite{Wildenthal71}.

We may therefore say that the monopole components of our realistic effective Hamiltonian account for the reduction in the separation of the $0i_{13/2}$ and $0h_{9/2}$ orbits. It is relevant to  point out that this reduction   has been interpreted as an effect of the neutron-proton tensor force \cite{Kay11,Otsuka05}, which makes us confident in the reliability of the tensor component of our interaction.

In summary, we have given here a shell-model description of $^{137}$Xe, focusing attention on the single-neutron structure recently  studied through the $^{136}$Xe($d,p$) reaction. Our shell-model effective interaction has been derived from the CD-Bonn $NN$ potential without using any adjustable parameter, in a way completely consistent with our previous studies of other nuclei in the $^{132}$Sn region.

We have obtained a very good agreement for both the measured excitation energies and the 
spectroscopic factors extracted from the data. In particular, the fragmentation of the single-neutron strength is well reproduced. These results make us confident in the predictive power of our effective interaction and provide motivation for a similar study of other nuclei around $^{132}$Sn which are within reach of ($ d,p$) transfer reactions with radioactive ion beams in inverse kinematics. This study is currently under way.

\begin{table}
\caption{Calculated energies and spectroscopic factors for states in $^{137}$Xe compared with those obtained from the experiment of Ref. \cite{Kay11} (see text for details).}
\begin{ruledtabular} 
\begin{tabular}{cccccc} 
\multicolumn{3} {c} {Expt.}  & \multicolumn{3} {c}  {Calc.} \\
\cline{1-3}     \cline{4-6}
$J^{\pi}$& E(MeV) & $C^{2}S$ & $J^{\pi}$ & E(MeV)  & $C^{2}S$ \\ 

\colrule $7/2^-$ & 0.000 & 0.94 & $7/2^{-}$ &  0.000 & 0.86\\ 
$3/2^-$ & 0.601 & 0.52 & $3/2^{-}$ &  0.728 & 0.57\\ 
$1/2^{-}$,$3/2^{-}$ &  0.986 & 0.35 & $1/2^-$ & 1.127 & 0.43 \\ 
$9/2^{-}$ &  1.218 & 0.43 & $9/2^-$ & 1.327 & 0.72 \\
$5/2^{-}$ &  1.303 & 0.22 & $5/2^-$ & 1.349 & 0.17\\ 
$5/2^{-}$,$7/2^{-}$ &  1.534 & 0.12 & $7/2^-$ & 1.589 & 0.05 \\
& & & $5/2^-$ & 1.666 & 0.04 \\
($9/2^{-}$) &  1.590 & 0.24 & $9/2^-$ & 1.584 & 0.01 \\
($13/2^{+}$) &  1.751 & 0.84 & $13/2^+$ & 2.082 & 0.75 \\
\end{tabular}
\end{ruledtabular}
\label{137xen}
\end{table}

\end{document}